# The impact of alloying elements on the precipitation stability and kinetics in iron based alloys: An atomistic study

G. Bonny[*,1], C. Domain[2], N. Castin[1], P. Olsson[3], L. Malerba[1,†]

[1] SCK•CEN, Nuclear Materials Science Institute, Boeretang 200, B-2400 Mol, Belgium.
[2] EdF-R&D, Département Matériaux et Mécanique des Composants, Les Renardières, F-77250 Moret sur Loing, France.
[3] KTH Royal Institute of Technology, Department of Physics, 106 91 Stockholm, Sweden.

## *Abstract*

Iron based industrial steels typically contain a large number of alloying elements, even so-called low alloyed steels. Under irradiation, these alloying elements form clusters that have a detrimental impact of the mechanical properties of the steel. The stability and formation mechanisms of such clusters are presently not fully understood. Therefore, in this work, we study the thermal stability and formation kinetics of small solute clusters in the bcc Fe matrix. We use density functional theory (DFT) to characterize the binding energy of vacancy/solute clusters containing Cr, Mn, Ni, Cu, Si and P, thereby exploring >700 different configurations. The constructed DFT data base is used to fit a cluster expansion (CE) for the vacancy-FeCrMnNiCuSiP system. In turn, the obtained CE is applied in atomistic kinetic Monte Carlo simulations to study the effect of Mn, Ni, Cr, Si and P on the precipitation formation in the FeCu alloy. We conclude that the addition of Mn and Ni delay the precipitation kinetics by an order of magnitude. The additional alloying with traces of P/Si further delays the kinetics by an additional order of magnitude. We found that Si plays an essential role in the formation of spatially mixed MnNiCuSi cluster.

[*] G. Bonny, corresponding author, gbonny@sckcen.be & giovanni.bonny@gmail.com
[†] Present affiliation: Centro de Investigaciones Energéticas, Medioambientales y Tecnológicas (CIEMAT), Avda. Complutense 40, 28040 Madrid, Spain

# 1. Introduction

Iron based industrial steels typically contain a large number of alloying elements. Besides C and N interstitial species, a range of substitutional alloying elements, such as Cr, Mn, Ni, Mo, V, and Si, are contained even in so-called low-alloyed steels, while Cu, P and S are often found as unwanted impurities. Alloying elements are added on purpose for metallurgical reasons, for example to facilitate the appearance of a certain microstructure (ferritic, bainitic, martensitic, austenitic), to provide solid solution strengthening, to induce suitable carbide formation and allow grain size control, or to improve corrosion resistance. Impurities are present due to the industrial production process [1].

In the nuclear industry, low alloyed bainitic steels are used to construct reactor pressure vessels (RPVs) [2, 3], while high-Cr ferritic-martensitic (FM) steels are envisaged to construct structural components in Generation IV reactors [4, 5] and for nuclear fusion applications [6]. Both RPV steels and FM steels harden and embrittle under irradiation at temperatures lower than 350-400°C [3, 5, 7]. In the case of RPV steels, Cu-rich precipitates are traditionally considered as the main cause for embrittlement [8], together with the so-called matrix damage, i.e., point-defect clusters [3]. In the case of high-Cr FM steels, Cr-rich precipitates are long known to be the cause of embrittlement under thermal ageing [9, 10] and to contribute also to radiation embrittlement [11], together with loops and voids [12]. In recent years, however, it has become clear that the minor alloying elements play a crucial important role in the microstructural evolution of the steel under irradiation and hence in its hardening and embrittlement [13-18].

In RPV steels, besides Cu-rich precipitates, MnNiSi-rich clusters are observed to form [19-21]. These clusters are claimed to correspond to (so-called) late blooming phases [22-24], although there is no consensus on whether they are actually thermodynamically stable phases or radiation-induced and stabilized features [15, 21, 23, 25].

In high-Cr FM steels, besides Cr-rich α' precipitates, NiMnSiP clusters are observed that do not correspond to any stable phase, but do have a significant effect on hardening and embrittlement [16-18, 26-28]. These clusters are often associated with dislocation loops and lines. It is therefore suspected that the mechanism at the origin of the formation of these solute clusters, irrespective of their nature as thermodynamically stable or radiation-stabilized features, is radiation-induced segregation on point-defect clusters [23, 29], driven by dragging of solutes by point defects, mainly vacancies [30, 31], but in the case of some elements also interstitials [30]. There is hence a need to develop suitable atomistic models that should ideally reproduce correctly both the thermodynamic and the kinetic properties of these complex alloys: the former provide the trend towards the formation of specific phases, while the latter account for detailed mechanisms of formation of clusters, including the role (if any) of the presence of point-defects that, under irradiation, are produced well above their equilibrium concentration.

Several atomistic modelling works [23, 25, 30, 32-66] address the problem of describing solute cluster formation and solute interaction with point defects. However, they are limited either by the chemical complexity included in the atomistic model or by the underlying physics grasped by the atomic interaction model. These models depend crucially on the reference DFT calculations, the representativeness of which remained so far limited by the cluster sizes addressed (generally only solute-solute or solute-point defect pairs).

With respect to high-Cr FM steels, DFT calculations addressed the interaction between Cr and point defects in iron [32, 33], more recently extending the study to triplets of Cr, Ni/Si/Mo/W atoms and single point defects [34-37]. However, a systematic study of the stability of solute clusters beyond pairs is so far missing. Interatomic potentials or other types of Hamiltonians that describe the binary FeCr [38-44], and the ternary FeCrW [37, 45] and FeCrNi [35] systems have been developed and widely applied to study the stability of Cr clusters near radiation defects [46, 47], as well as the precipitation kinetics in high-Cr FM steels under thermal ageing [42, 43, 48-51]. The physics includes many-body interactions via the two-band model (2BM) [38], concentration dependent model (CDM) [39] and (concentration dependent) cluster expansion (CE) [43], but the chemical complexity (up to ternary alloys) remains limited.

With respect to RPV steels, DFT calculations addressed the interaction between solutes (Cu/Ni/Mn/Si/P) and point defects [25, 30, 52-54]. However, here too a systematic study of the stability of solute clusters beyond pairs is missing. Interatomic potentials have been developed for the FeCu [44, 55, 56], FeCuNi [57], FeCuNiMn [25] systems and rigid lattice pair interaction models for the FeCu [58] and FeCuNiMnSiP systems [23, 59-61]. These models have been widely applied to study the thermal stability of solute clusters near defects [62-65] and precipitation kinetics in RPV steels under irradiation [23, 29, 59-61] and during thermal ageing [54, 58, 66]. Even though the interatomic potentials, which describe the many-body interactions via the embedded atom method (EAM) [67], were pushed to include a quaternary alloy, their chemical complexity remains insufficient to deal with fully realistic systems. The pair interaction models, on the other hand, include up to six chemical elements, but lack many-body interactions.

In the present work, we take a step forward towards an all-including atomistic model by removing most of the shortcomings of previous work. We performed a systematic DFT study of solute and solute-vacancy clusters in the bcc Fe matrix, ranging from pairs up to quintuplets for vacancies (v) bound to Cr, Mn, Ni, Cu, Si, and P atoms, i.e., the six species that are typically found experimentally in solute clusters formed under irradiation in iron based alloys. More than 700 cluster configurations are explored. This data provide insight into the possibility of precipitation and co-precipitation of different alloying elements in iron. We also improve the present state-of-the-art in the description of the chemical complexity of steels by fitting a CE that includes many-body effects (up to triplets) for the 8-component FeCrMnNiCuSiP-vacancy system, valid in the Fe-rich limit. This CE is then applied in an atomistic kinetic Monte Carlo (AKMC) scheme to simulate thermal ageing, i.e., to study the stability and kinetics of formation of complex solute clusters at relatively high temperature, focusing on the vacancy-driven mechanism. Specifically, the CE allows the effects of the minor alloying elements Cr, Mn, Ni, Si, P, on the Cu precipitation kinetics and stability to be investigated in a range of temperatures of practical interest, by isolating the effect of each element and in general covering a set of conditions that would be very long and expensive to explore experimentally.

The paper is organized as follows. Following this introduction, in section 2, we provide a brief description of the methods applied for the calculation of cluster stabilities via DFT, the CE development and the simulation of thermal ageing via AKMC. In section 3.1, we present and discuss the results of the DFT calculations in terms of the impact of the different elements on solute cluster stabilities. In section 3.2, the CE is presented and validated against CALPHAD and DFT data. In section 3.3, we present the results of the simulated thermal ageing and discuss the impact of the several elements on the precipitation kinetics and cluster morphology. The paper is finalized by conclusive remarks.

## 2. Methods

### 2.1. Solute cluster stability

The binding energy of solute and solute-vacancy (v) clusters in a bcc Fe matrix was calculated by DFT. The total binding energy, $E_b$, of a cluster $\{S_i\}_N$, with $S_i =$ v, Cr, Mn, Ni, Cu, Si, P, consisting of $N$ solutes/vacancies is given as,

$$E_{\text{b}}(\{S_i\}_N) = \sum_i E(S_i) - [E(\{S_i\}_N) + (N-1)E_{\text{ref}}], \tag{1}$$

with $E_{\text{ref}}$ the total energy of the reference system (bcc Fe), $E(S_i)$ the total energy of the system containing a solute/vacancy $S_i$ and $E(\{S_i\}_N)$ the total energy of the system containing the cluster $\{S_i\}_N$. Given this definition, positive values of $E_{\text{b}}$ denote attraction. In turn, the incremental binding energy, $E_{\text{b}}^{\text{inc}}$, of a solute/vacancy $S_i$ to a cluster $\{S_i\}_N$ is obtained as,

$$E_{\text{b}}^{\text{inc}}(S_i) = E_{\text{b}}(\{S_i\}_N) - E_{\text{b}}(\{S_i\}_{N-1}). \tag{2}$$

The energy of the configurations that enter equation 1 were calculated using the Vienna ab initio simulation package (VASP) [68, 69], using the Projector Augmented Wave (PAW) method [70, 71] in the Generalized Gradient Approximation (GGA), with Perdew-Wang (PW91) exchange-correlation functional [72], and Vosko-Wilk-Nusair interpolation [73]. Spin polarized potentials with 8, 6, 7, 10, 11, 4 and 5 valence electrons for Fe, Cr, Mn, Ni, Cu, Si and P were used, respectively. The Methfessel-Paxton method with finite temperature smearing was applied (smearing width 0.3 eV). To obtain convergence, the plane-wave cut-off energy was 300 eV and a 3×3×3 k-points grid following the Monkhorst-Pack scheme proved sufficient. All calculations were performed in periodic bcc supercells containing 128 atoms with a fixed lattice parameter set to the theoretical value of Fe (2.831 Å). With the present settings, a convergence error of ~10 meV on the total binding energy (see equation 1) can be estimated.

### 2.2. Cluster expansion

A CE was fitted to the 8-component vFeCrMnNiCuSiP system. Cluster interactions up to triplets (many-body interactions) were included. The CE is defined by a total of 260 clusters and cluster interactions: 1 constant term, 7 point terms, 28 first (1nn) and 28 second (2nn) nearest neighbor pair terms, and 196 triangle terms. A full description of the CE, including form of the cluster functions, is provided in the Appendix. The optimized interaction coefficients are provided in the supplementary material.

The CE was fitted using a similar methodology as in [45]. The reference quantities are: cohesive energy of the pure species; binding energy of solute-solute, vacancy-vacancy and vacancy-solute clusters in the bcc Fe matrix; formation energy of a vacancy in the bulk of each element; substitutional energy of all elements in the bulk of one-another; and the experimental solubility limit of each element in bcc Fe. However, the different quantities are fitted with different weights, depending on their importance for the physics of the problem, i.e., the solute/vacancy cluster stability in the bcc Fe matrix.

The highest weight was given to the binding energy of v-solute pairs and the solubility limit of all solutes in the bcc Fe matrix. Given the targeted clustering behavior under irradiation and thermal ageing, these quantities are essential. The v-solute binding energy is also important to reproduce v-solute dragging in a certain temperature range (see for example [56]). The solubility limit of each element in the bcc Fe matrix defines the thermodynamic driving force to form solute clusters. Even though such information is already implicitly included in the DFT calculated binding energy of solute clusters, additional fine-tuning is necessary. In particular, the DFT computed energetics is not always consistent with the experimental solubility limit, as discussed for example in [44].

Medium weight was given to the binding energy of vacancy/solute clusters in the bcc Fe matrix and the vacancy formation energy in the bulk of each element. The binding energy of vacancy/solute clusters provides a measure of their stability, which is essential when studying materials under irradiation conditions. The vacancy formation energy is added to avoid artificial trapping of a vacancy in large solute clusters. Once again, even though this information is implicitly contained in the binding energy of vacancy-solute clusters, additional fine-tuning is necessary.

Low weight was given to the formation energy of all solutes in one-another and to the cohesive energy of each element. The formation energy is added to avoid unphysical values and thus the possible formation of unphysical phases/clusters. Once more, such information is implicitly contained in the binding energy of solute clusters, but additional fine-tuning is necessary. Finally, the cohesive energy of each element is added for completeness, but in fact, it does not play any role in the AKMC simulations.

The fitting of the solubility limit of the different solutes in the bcc Fe matrix was controlled by modification of the solute formation energy in bcc Fe, $E_f(S \text{ in Fe})$. In the first iteration of the fitting procedure, we relate $E_f(S \text{ in Fe})$ to the solubility of a solute $S$ in bcc Fe, $C_{\text{Fe}}^S$, via Raoult's activity law for ideal solutions,

$$C_{\text{Fe}}^S = \exp\left(\frac{-E_f(S \text{ in Fe})}{k_B T}\right), \tag{3}$$

with $T$ the absolute temperature and $k_B$ Boltzmann's constant. After the initial guess for $E_f(S \text{ in Fe})$, a trial CE is constructed. This CE is used in Monte Carlo simulations employing the transmutation (semi-grand canonical) ensemble to construct the solubility limit curve, following the strategy developed in [74]. By comparing the obtained $C_{\text{Fe}}^S$ to the target (CALPHAD calculated) value, $E_f(S \text{ in Fe})$ is modified accordingly and a new CE is constructed. In this way, $E_f(S \text{ in Fe})$ is fine-tuned by trial and error to reproduce the experimental solubility limit of every considered element in the bcc Fe matrix.

### 2.3. Simulated thermal ageing

We simulate thermal ageing processes by AKMC methods. Initially, a bcc random alloy with fixed alloy composition is created in a simulation volume where atoms occupy the nodes of a rigid lattice. During the AKMC simulation, the (solute) atoms are redistributed by exchanging position with a diffusing single vacancy, which corresponds to the mechanism that is active under thermal ageing conditions. At each AKMC step, the transition probability of the vacancy from its initial position $i$ to one of its neighboring sites (8 for the bcc lattice) $j$, $\Gamma_{i \to j}$, is given as,

$$\Gamma_{i \to j} = \nu_0 \exp\left(\frac{-E_m^{i \to j}}{k_B T}\right). \quad (4)$$

Here $E_m^{i \to j}$ is the migration energy for the vacancy to move from position $i$ to $j$; and $\nu_0$ the attempt frequency, here taken equal to the typical value of $6 \times 10^{12}$ s$^{-1}$.

The simulations were performed in cubic bcc crystals containing 128,000 atoms with a single vacancy. The Monte Carlo time, $t_{\text{MC}}$, was estimated via the mean residence time algorithm [75]. However, during the simulations, the vacancy can get trapped in solute clusters (in our case Cu-rich clusters). In this case, the time spent by the vacancy in the cluster does not contribute to the physical evolution of the system. To account for this, we followed the method proposed by Soisson *et al.* [58], accounting for the fraction $f_v$ of MC time when the vacancy is found in a locally Cu-free configuration. In addition, the resulting Monte Carlo time still differs from the real time, $t_{\text{real}}$, due to the difference in vacancy concentration between simulation and real alloy. To obtain $t_{\text{real}}$, $t_{\text{MC}}$ must be rescaled, as

$$t_{\text{real}} = \frac{\frac{f_v}{N_{\text{box}}}}{\exp\left(-\frac{E_f^v}{k_B T}\right)\exp\left(-\frac{S_f^v}{k_B}\right)} \ t_{\text{MC}} = \frac{f_v}{3N_{\text{box}}\exp\left(-E_f^v/k_B T\right)} \ t_{\text{MC}}. \tag{5}$$

Here, the numerator gives the vacancy concentration in the simulation box, while the denominator estimates the thermal vacancy concentration in the real alloy, with $N_{\text{box}}$ the number of atoms in the simulation box and $E_f^V$ the vacancy formation energy in the alloy (the factor 3 is a numerical estimate of the entropic contribution [76]).

The precise value of $E_m^{i \to j}$ (see equation 4) depends on the local chemical environment (LCE) around the migrating vacancy and estimating it correctly is essential to predict a realistic evolution of the system. In the present work, we use the Kang-Weinberg (KW) approximation to estimate the LCE dependent $E_m^{i \to j}$, which has been successfully applied to simulate thermal ageing [45, 48-50, 77]. Within this approximation, $E_m^{i \to j}$ is given as,

$$E_m^{i \to j} = E_{m0}^j + \frac{\Delta E_{j-i}}{2}, \tag{6}$$

where $\Delta E_{j-i}$ is the total energy change after and before the jump, and $E_{m0}^j$ is the excess migration energy, depending on the atom type at site $j$. In this way, $E_{m0}^j$ is symmetric and provides a measure for the kinetics in the random alloy, while $\frac{\Delta E_{j-i}}{2}$ provides the driving force towards the thermodynamic limit. As a first approximation, $E_{m0}^j$ is taken as the vacancy exchange barrier of the solute type at site $j$ in an otherwise pure bcc Fe matrix.

## 3. Results and discussion

### 3.1. Energetics of solute clusters in iron

The total and incremental binding energy for all studied configurations (>700) are tabulated in the supplementary material. In Figure 1, the normalized binding energy, $E_b/N$, as calculated with DFT for 1nn and 2nn pairs, triplets and quartets are graphically summarized. The topology of the clusters is shown in the figure inset and includes all possible configurations for 1nn and 2nn pairs (2×28), {1nn,1nn,2nn} triangles (196) and tetrahedra (406). The figure provides indications regarding the stability of the different solute/vacancy clusters. For an accurate analysis, however, $E_b^{inc}$, rather than $E_b/N$, should be analyzed, as discussed below and provided in detail in the supplementary material. The values for $E_b^{inc}$ can be readily implemented in cluster dynamics and object kinetic Monte Carlo codes and used to define the kinetics of dissociation of clusters.

The data in Figure 1 is consistent with available data in the literature: CrCr and vCr pairs [32]; solute-solute and v-solute pairs [30, 34]; mixed MnNi, NiCu and MnCu pairs [25]; and mixed NiCr pairs [35]; Cr triplets and quartets [32]; and vNiCr triplets [35].

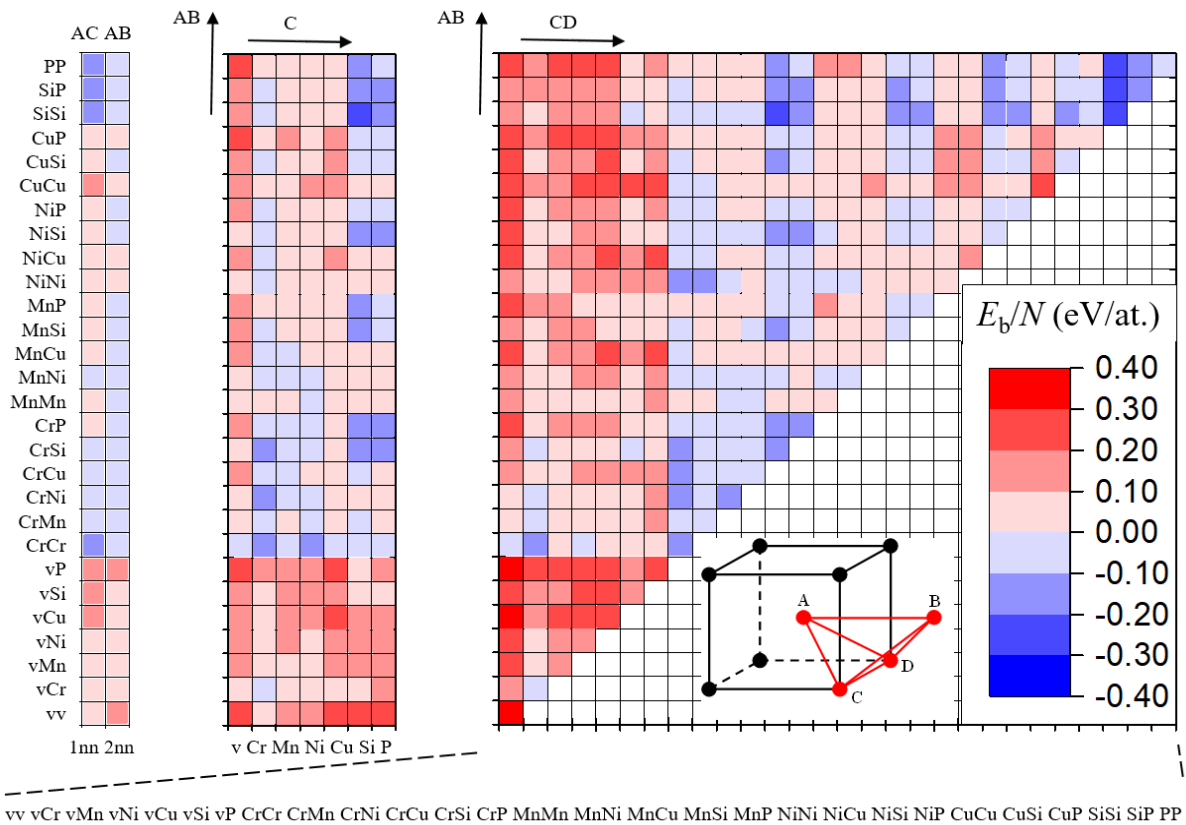

Figure 1 – Graphical summary of the DFT values for $E_b/N$ for all possible cluster configurations including 1nn and 2nn pairs, triplets and quartets. The topology of the clusters is provided in the inset of the figure and $E_b/N$ is given in eV/atom.

In the following, the stability of solute/vacancy clusters in a bcc Fe matrix are discussed in terms of their ability to mix in the same clusters. The analysis is based on $E_b^{inc}$ obtained from the DFT data presented in Figure 1. In with $E_\text{b} \leq -0.05$ eV are considered repulsive. In the table, the cells corresponding to binding pairs are color-coded green, neutral pairs white and repulsive pairs red. For triplets and quartets, we define the stability as follows: $\min[E_b^\text{incr}(S_i)] \geq 0.05$ eV corresponds to a binding cluster; $\min[E_b^\text{incr}(S_i)] \leq -0.05$ eV corresponds to a repulsive cluster; and $-0.05\,\text{eV} < \min[E_b^\text{incr}(S_i)] < 0.05$ eV corresponds to a neutral cluster. In the table, binding triplets/quartets are written in bold, while neutral triplets/quartets are written in italic.

Table 1, all possible combinations of 1nn (cells in Table 1 above the diagonal) and 2nn (cells in Table 1 below the diagonal) vacancy/solute pairs are enumerated. Pairs with $E_b \geq 0.05$ eV are considered binding; pairs with $-0.05$ eV $< E_b < 0.05$ eV are considered neutral; and pairs

Clearly, species that form binding or neutral pairs can mix into larger clusters. For repulsive pairs, however, the situation is less straightforward. Their combination with other pairs might result in stable or neutral clusters, and thus mixing. Therefore, in Table 1, we add in bold all triplets that contain a neutral or repulsive pair in the corresponding cell of the repulsive and/or neutral pair; and in italic we add all the triplets that contain a repulsive pair but are neutral in the corresponding cell of the repulsive pair. For quartets the same is done, with the additional condition that for neutral quartets, the cluster cannot contain a neutral or binding triplet.

In this way, we identify all non-trivial neutral and binding triplets and quartets. Thus, by the addition of certain elements to repulsive pairs, the larger resulting cluster can be stabilized, allowing the elements to be mixed into a single cluster. As a consequence, we see that already starting from triplets, all elements can be combined in a cluster, and the same holds for quartets. Thus, the fact that individual pairs are repulsive, does not necessarily mean that the solutes cannot combine into a larger solute cluster. In particular, we conclude that while CrCr, SiSi, PP and SiP pairs repel, they can be stabilized in larger clusters (starting from triplets) by the addition of other alloying elements.

Table 1 – Qualitative representation of the ability of mixing of different species into a single cluster. Cells corresponding to binding, neutral and repulsive pairs are colored in green, white and red, respectively. Binding triplets/quartets are written in bold, while neutral triplets/quartets are written in italic. The cells in the table above the diagonal represent 1nn pairs while the cells below it represent 2nn pairs.

| 2nn\1nn | v | Cr | Mn | Ni | Cu | Si | P |
|---|---|---|---|---|---|---|---|
| v | | | | | | | |
| Cr | v v Cr | *Mn Cr Cr*<br>*P Cr Cr*<br>**v Mn Cr Cr** | v Mn Cr Cr | *Cr Ni Mn Ni*<br>*Cr P Mn Ni* | | Cr Mn Si P | Cr v P<br>Cr Mn Si P |
| Mn | | Cr Mn Si P | **Mn Mn Mn**<br>**Mn Mn Si**<br>**v Mn Mn**<br>**Cu Mn Mn** | **Mn v Ni**<br>**Ni v Mn**<br>**Mn Ni Mn Cu**<br>**Mn Ni Mn P** | | | |

|    |    |                                    | Si Mn Mn<br>P Mn Mn<br>Mn Ni Mn Cu<br>Mn Ni Mn P<br>Mn Cu Mn Cu |                                           |                                      |                                                                    |                                             |
|----|----|------------------------------------|------------------------------------------------------------------|-------------------------------------------|--------------------------------------|--------------------------------------------------------------------|---------------------------------------------|
| Ni |    | *v Cr Ni*<br>*P Cr Ni*<br>*Cr Ni Mn Ni* | *v Mn Ni*<br>*Si Mn Ni*<br>*Cr Ni Mn Ni*<br>*Cr P Mn Ni*<br>*Mn Ni Mn Ni*<br>**Mn Ni Mn Cu**<br>**Mn Ni Mn P**<br>**Mn Ni Si Si**<br>**Mn Ni Si P** | Ni v Ni<br>v Ni Ni<br>Cu Ni Ni<br>Si Ni Ni<br>P Ni Ni |                                      |                                                                    |                                             |
| Cu |    |                                    | *v Mn Cu*<br>*Cr Mn Cu*<br>*Si Mn Cu*<br>*P Mn Cu*<br>**Mn Ni Mn Cu**<br>**Mn Cu Mn Cu**<br>**Mn Cu Si Si** | v Ni Cu<br>Cu Ni Cu<br>Si Ni Cu          |                                      |                                                                    |                                             |
| Si |    | *v Cr Si*                          | Mn Mn Si                                                         | *v Ni Si*<br>*Ni Ni Si*<br>*Cu Ni Si*     | *v Cu Si*<br>*Ni Cu Si*<br>*Cu Cu Si* | v Si Si<br>Mn Si Si<br>Ni Si Si<br>Cu Si Si<br>**Mn Ni Si Si**<br>**Mn Cu Si Si** |                                             |
| P  |    | *v Cr P*<br>*Cr P Mn Ni*            | *v Mn P*<br>**Mn Ni Mn P**                                        | *v Ni P*<br>*Mn Ni P*<br>*Ni Ni P*<br>**Cu Ni P** |                                      | v Si P<br>Mn Si P<br>Ni Si P<br>**Cu Si P**<br>**Cr Mn Si P**<br>**Mn Ni Si P** | P v P<br>v P P<br>Cr P P<br>Mn P P<br>Ni P P<br>Cu P P |

## 3.2. Cluster expansion

In this section, we assess the validity of the CE by comparing its results to the CALPHAD calculated $C_{\mathrm{Fe}}^{S}$, $E_\mathrm{b}$ of v-solute pairs, $E_\mathrm{b}/N$ of vacancy/solute clusters, solute formation energy and cohesive energy.

In Table 2, we compare the solubility limit of all the elements in the bcc Fe matrix as obtained from Monte Carlo simulations following [74] with the values obtained from the SSOL2 thermodynamic data base [78], based on experimental data. For Cr, in addition, the value based on a recent revision was added [79, 80]. Except for Cu and P, all values were obtained for $T = 600$ K, which corresponds to typical current nuclear reactor operating conditions. For Cu and P, we selected $T$=1000 K because of their negligible solubility at 600 K. As can be seen, the agreement between the $C_{\mathrm{Fe}}^{S}$ obtained by CE and the reference data is excellent.

In addition, in Figure *2* we compare the solubility limit versus temperature predicted by the CE with the CALPHAD results obtained using the SSOL2 thermodynamic database. At low temperature, the CE somewhat overestimates the CALPHAD solubility of P, Cu and Ni and underestimates the CALPHAD solubility of Si. We note that the deviation of the Ni and Mn solubility at 650 K and 500 K from the CALPHAD one is the result of the bcc to fcc transition in Fe, which cannot be captured in the rigid lattice model. Also, we note that because of the low content of Si in standard steels, the underestimation of the Si solubility is irrelevant for the applications targeted in this work. Overall, given the simplicity of the model, agreement is satisfactory.

For the FeCr system, the two limiting cases for the solubility limit are presented. The parameterization by Andersson et al. [81] is based on high temperature data while the one by Bonny et al. [79, 80] accounts for the change of sign and curvature of the mixing enthalpy as, predicted by DFT [82]. To reproduce the complex shape of the mixing enthalpy as predicted by DFT, a cluster expansion including cluster sizes up to quintuplets is necessary [42]. For the present applications, these low temperature effects were not introduced in the CE.

Table 2 – Comparison between the solubility limit at 600 K (1000 K for Cu and P) obtained from the SSOL2 database with the values obtained from Monte Carlo simulations using the CE.

| $C_{Fe}^S$ / S | Cr* | Mn | Ni | Cu | Si | P |
|---|---|---|---|---|---|---|
| SSOL2 (at. %) | 5-8 | 3.7 | 2.9 | 0.7 | 6.2 | 1.3 |
| CE (at. %) | 6.7±0.2 | 3.6±0.3 | 2.8±0.3 | 0.7±0.1 | 6.0±0.2 | 1.4±0.1 |

* Limiting values based on the parameterizations by Andersson *et al.* [81] and Bonny *et al.* [79, 80].

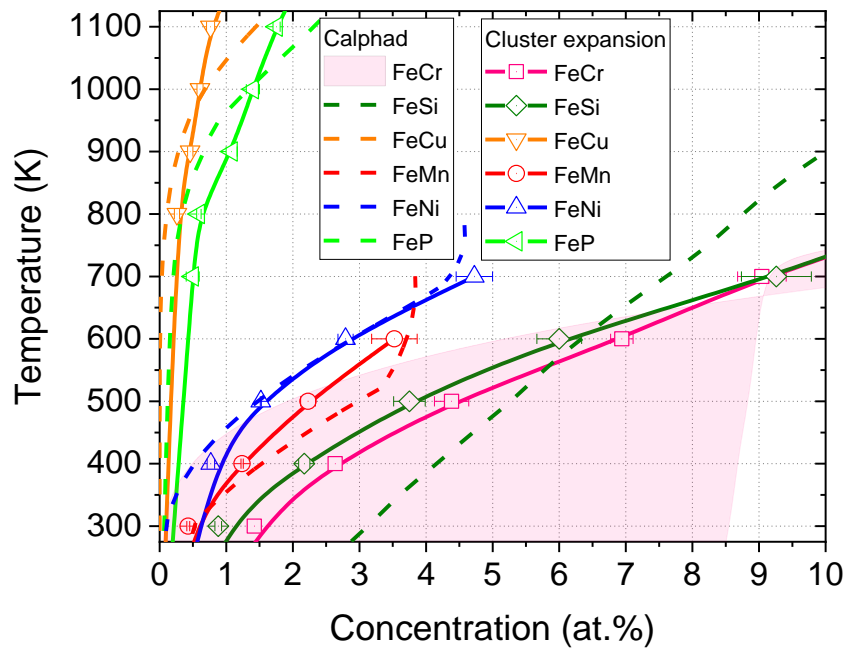

Figure 2 – Comparison between the solubility limit obtained from the CE via MMC simulations with the ones obtained from the CALPHAD SSOL2 database.

In Table 3, a comparison between DFT and the CE for the $E_b$ of v-$S$ pairs in bcc Fe is presented. For all elements, the binding energy is well reproduced, both qualitatively and quantitatively. All v-$S$ pairs are attractive, except for neutral interaction in the case of v-Cr pairs. In addition, the correct minimum energy configurations are reproduced: 2nn pairs for v-v and v-Ni pairs, 1nn pairs for v-Mn, v-Cu, v-Si and v-P pairs and a degenerate state for v-Cr pairs.

Table 3 – Comparison between DFT and the CE concerning the $E_b$ of v-$S$ pairs in bcc Fe. All values are presented in eV and values in parentheses were obtained from the CE.

|     | v | Cr | Mn | Ni | Cu | Si | P |
| --- | --- | --- | --- | --- | --- | --- | --- |
| 1nn | 0.15 (0.19) | 0.06 (-0.02) | 0.17 (0.20) | 0.12 (0.13) | 0.27 (0.28) | 0.31 (0.29) | 0.38 (0.36) |
| 2nn | 0.21 (0.26) | 0.02 (-0.02) | 0.11 (0.10) | 0.20 (0.21) | 0.16 (0.18) | 0.11 (0.10) | 0.27 (0.23) |

In Figure 3, the normalized binding energy, $E_b/N$, of pure solute and pure vacancy clusters as a function of cluster size calculated by both DFT and the CE is presented. For v, Mn, Ni and Cu, the agreement between DFT and the CE is excellent. For Cr, Si and P clusters, however, the discrepancy is large. For the latter elements, DFT predicts repulsive interaction whereas the CE gives neutral interaction. The large discrepancy is a consequence of the negative heat of mixing of Cr, Si and P in bcc Fe, as derived from the SSOL2 data base and DFT [82]. Within the framework of many-body interactions up to triplets, however, it is impossible to satisfy both the CALPHAD calculated solubility and repulsive $E_b/N$. To satisfy both $C_{Fe}^S$ and $E_b/N$, many-body interactions up to at least quintuplets need to be included (see [50]), which is beyond the scope of the present work. Given the target application, i.e., the formation of solute clusters, priority was given to a correct reproduction of $C_{Fe}^S$. As is shown in the following, this shortcoming has limited consequences on the stability of mixed vacancy/solute clusters.

We emphasize, in any case, that even though only DFT data up to triplets were included in the fit, still the extrapolation towards larger clusters is excellent (except for Cr, Si and P). This indicates that triplet many-body interactions in the CE (except for Cr, Si and P) suffice to catch the DFT logic.

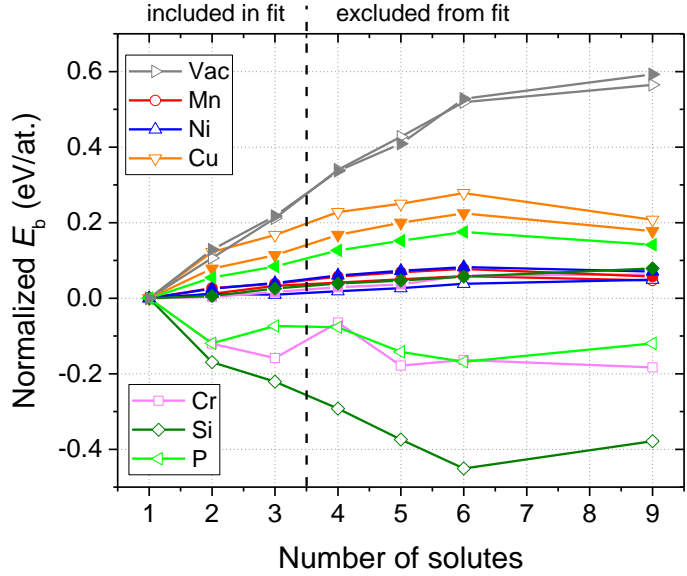

Figure 3 – Normalized binding energy of pure solute and pure vacancy clusters as a function of cluster size calculated by both DFT and the CE. Empty symbols indicate DFT data and filled symbols indicate CE data.

Given the inconsistencies between DFT and the CE for Cr, Si and P discussed above, in the following, we shall compare the data for clusters containing Cr, Si or P separately from the others. In Figure 4, $E_b/N$ as calculated by DFT and the interaction model for all considered clusters not containing Cr, Si or P are presented. The average error between the DFT and CE data sets is 21 meV, with a correlation factor of $R^2=0.92$. For the data included in the fit, the average error is 14 meV with a correlation factor of $R^2=0.94$, while for the data not included in the fit, the average error is 26 meV with a correlation factor of $R^2=0.92$.

For comparison, we have also included the same plot (Figure 4b) for the pair interaction model by Vincent et al [59]. For that model, the average error between DFT and model is 61 meV and the correlation factor is $R^2 = 0.67$. The low correlation factor indicates that the pair interaction model did not capture the DFT logic. This can be linked to the absence of many-body interaction terms in the pair interaction model.

Thus, for our CE, the average error for both reference data set and fitting data set is similar, and for all data sets the correlation factor is high. The fact that a fit up to triplets can extrapolate data up to at least nonuplets illustrates the transferability of the CE. On average, the bias between the DFT and CE data sets is small. The CE overestimates the DFT data set by 9 meV on average. This small bias is possibly the result of inconsistencies between the SSOL2 database and the used DFT method.

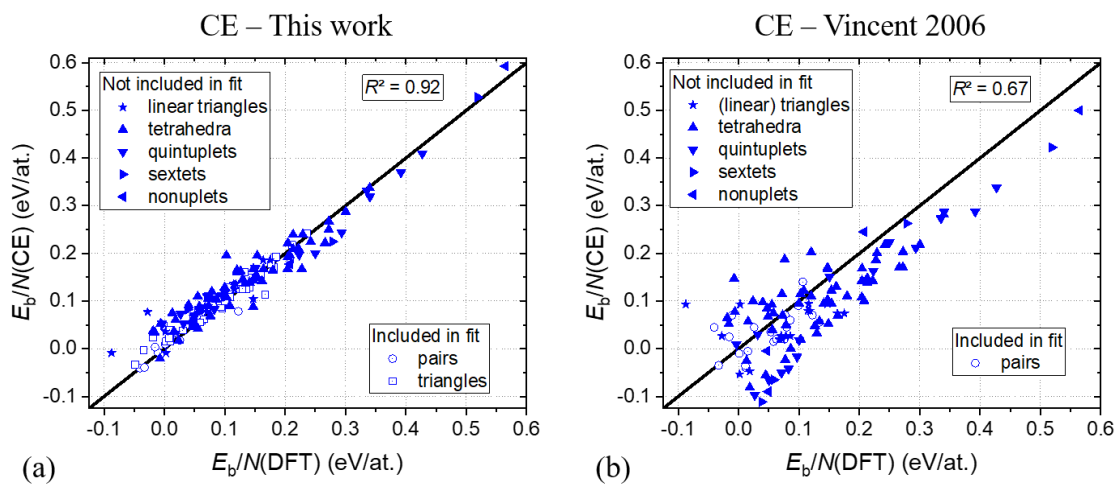

Figure 4 – Comparison of the normalized binding energy calculated by DFT and the CE for the different solute clusters that do not contain Cr, Si or P.

In Figure 5, $E_b/N$ as calculated by DFT and the CE for all considered clusters containing Cr, Si or P are presented. The data sets were separated following their Cr, Si or P content in the clusters, i.e., clusters containing ≤ 25%, between 25-75% and ≥ 75% Cr, Si or P. With a correlation factor $R^2=0.11$, there is no correlation between the DFT and CE data sets for clusters containing ≥ 75% Cr, Si or P. The average error between both data sets is 132 meV and the CE overestimates the DFT data by about the same amount. For data sets below 75%, however, there is a strong correlation. For concentrations between 25-75% we obtain a correlation factor $R^2=0.82$, and $R^2=0.97$ for concentrations ≤ 25%. The average error between DFT and CE data sets is 34 meV (12 meV overestimation bias by the CE) and 12 meV (no bias) for concentration between 25-75% and ≤ 25%, respectively. Although the average error is higher than for cluster that do not contain Cr, Si or P, there is a strong correlation between the DFT and CE data sets, which indicates that for the latter compositions the DFT logic is followed by the CE. Moreover, when considering the data that was not included in the fit, we obtain the correlation factor $R^2=0.91$ with average error 61 meV and $R^2=0.95$ with average error 39 meV for clusters containing between 25-75% and ≤ 25%, respectively. This means that, for clusters containing less than 75% Cr, Si or P, the CE catches the DFT logic and that many-body interactions in the CE up to triplets suffice to predict larger configurations that were not included in the fit. We can thus also conclude that, so long as vacancy/solute clusters do not contain more than 75% Cr, Si or P, the CE is a reliable extrapolation tool for the DFT data.

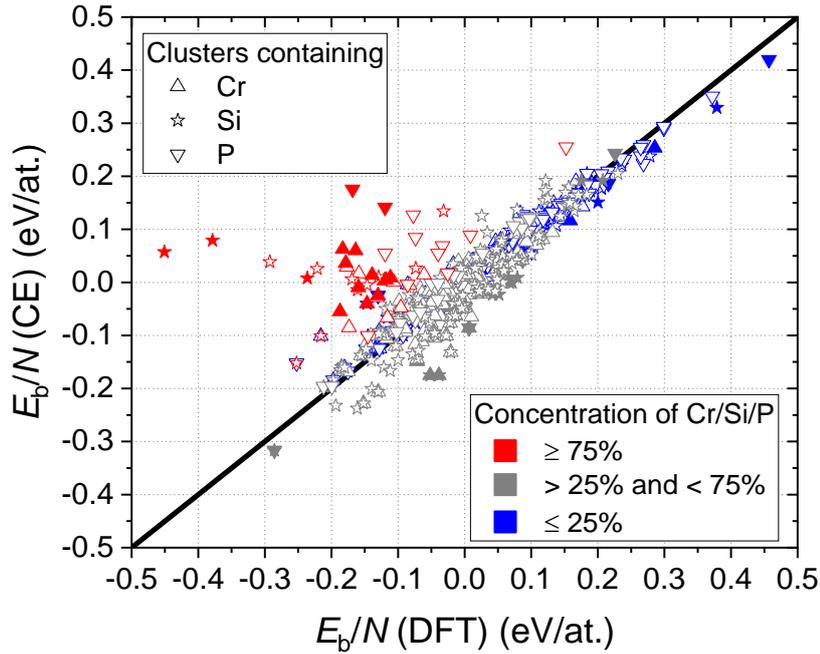

Figure 5 – Comparison of the normalized binding energy calculated by DFT and the CE for different solute clusters that do contain Cr, Si or P. Data represented by open symbols were included in the fit, while data represented by filled symbols were not included in the fit.

To avoid possible artificial trapping of vacancies in solute clusters, the vacancy formation energy in the bulk of each element was included in the fit. Since the CE is a rigid lattice model based on the bcc crystalline structure, we fitted the vacancy formation energy in the bulk of each element in a bcc crystal, even when this is not the ground state of the element, as is the case for Mn, Ni, Cu, Si and P. Since the DFT calculation of the vacancy formation energy in a crystal that is not in the ground state is not always possible, we used an empirical regression to estimate the vacancy formation energy. The target vacancy formation energy was taken based on the regression $T_m/1000$, with $T_m$ the melting temperature. The latter regression provides a reasonable estimate, as compared to, for example, the DFT values for Fe (1.9-2.2 eV) and Cr (2.5-2.8 eV) [32]. This way of proceeding has been considered sufficient because what matters is not that the vacancy formation energy in bcc pure elements is correct in absolute terms, but only that unphysically low values that may stem as artefacts of the fitting procedure should be avoided.

Table 4 – Comparison of the target vacancy formation energy with the one obtained by the CE. The target vacancy formation energy is obtained from the regression $T_m/1000$, with $T_m$ taken from [83]. All energies are reported in eV.

|                     | Fe   | Cr   | Mn   | Ni   | Cu   | Si   | P    |
|---------------------|------|------|------|------|------|------|------|
| Target ($T_m/1000$) | 1.81 | 2.13 | 1.52 | 1.73 | 1.36 | 1.69 | 0.32 |
| CE                  | 1.78 | 2.14 | 1.52 | 1.73 | 1.36 | 1.70 | 0.33 |

To obtain reasonable mixing properties between the different elements, we also fitted the formation energy of element $Y$ in element $X$, $E_f(Y \text{ in } X)$. These properties are implicitly included via the DFT binding energy, but need refinement to avoid unphysical values (without this refinement values of the order 10 eV are not uncommon). The values for $E_f(Y \text{ in } X)$ were derived from the mixing enthalpy obtained from the SSOL2 database at 300 K (the lower temperature limit to use SSOL2) as,

$$E_f(Y \text{ in } X) = \left.\frac{dH_{mix}}{dC_Y}\right|_{C_Y=0}. \tag{7}$$

A comparison between the target values and CE for $E_f(Y \text{ in } X)$ is given in Table 5. Agreement in the first row and column of the table is poor. The first row represents $E_f(Y \text{ in Fe})$, which was fitted directly to reproduce the solubility limit. The first column represents $E_f(\text{Fe in } X)$, which is also controlled by the fit to the solubility limit; i.e., given the limited degrees of freedom we chose the mixing enthalpy to be as symmetric as possible. For all other combinations, the agreement is reasonable, given the low weight given to this quantity in the fit.

Table 5 – Formation energy of $Y$ in $X$. All values are in eV and the values corresponding to the CE are given in parentheses.

| $X\backslash Y$ | Fe      | Cr      | Mn      | Ni      | Cu     | Si      | P       |
|-----------------|---------|---------|---------|---------|--------|---------|---------|
| Fe              |         | 0.21    | 0.09    | 0.02    | 0.45   | -1.85   | -1.58   |
|                 |         | (0.21)  | (0.24)  | (0.26)  | (0.65) | (0.21)  | (0.55)  |
| Cr              | 0.31    |         | -0.37   | 0.53    | 0.96   | -1.51   | 0.00    |
|                 | (0.30)  |         | (0.12)  | (1.13)  | (1.72) | (-1.11) | (0.41)  |
| Mn              | -0.10   | -0.37   |         | -0.66   | 0.00   | -1.28   | 0.00    |
|                 | (0.23)  | (-0.44) |         | (-0.61) | (0.04) | (-1.41) | (-0.20) |
| Ni              | -0.32   | 0.06    | -0.51   |         | 0.17   | 0.02    | -1.39   |
|                 | (0.27)  | (0.04)  | (-0.42) |         | (0.23) | (-0.18) | (-1.35) |
| Cu              | 0.57    | 0.83    | 0.02    | 0.06    |        | -0.38   | -1.36   |
|                 | (0.57)  | (0.63)  | (0.10)  | (0.02)  |        | (-0.55) | (-1.33) |
| Si | 0.00 (0.22) | 0.00 (0.78) | 0.00 (1.02) | 0.00 (0.97) | 0.00 (0.99) | | 0.00 (0.60) |
| P | 0.00 (0.55) | 0.00 (0.66) | 0.00 (0.95) | 0.00 (0.74) | 0.00 (0.67) | 0.00 (0.24) | |

Finally, we present the cohesive energy, $E_c$, of the pure species as reproduced by the CE on a bcc lattice. The target value for each element was determined for the bcc lattice as the sum of $E_c$ of the true ground state (GS) and the energy difference, $\Delta E$, between the true ground state and the bcc phase. The $E_c$ for the true ground state is taken from [83], while $\Delta E$ is calculated from the Calphad SGTE database at 300 K [84]. In Table 6, the target data is summarized. For all elements, the resulting $E_c$ for the bcc phase, $E_c(A2)$, is identical to the predictions of the CE.

Table 6 – Cohesive energy for the different elements. The cohesive energy in the ground state, $E_c(GS)$, was taken from [83], the energy difference between true ground state and bcc phase, $\Delta E$, was taken from [84]. All values are reported in eV.

| | V | Fe | Cr | Mn | Ni | Cu | Si | P |
|---|---|---|---|---|---|---|---|---|
| GS | Vacuum | A2 | A2 | A12 | A1 | A1 | A4 | white P |
| $E_c(GS)$ | 0.00 | 4.28 | 4.10 | 2.92 | 4.44 | 3.49 | 4.63 | 3.43 |
| $\Delta E$ | 0.00 | 0.00 | 0.00 | 0.04 | 0.08 | 0.04 | 0.51 | 0.28 |
| $E_c(A2)$ | 0.00 | 4.28 | 4.10 | 2.88 | 4.36 | 3.45 | 4.12 | 3.15 |

### 3.3. Simulated thermal ageing

In this section, we present the results of simulating thermal ageing experiments at 638 K for a composition corresponding to a high-Cu/high-Ni RPV steel (see Table 7), taken from the experimental work by Styman *et al.* [85]. Direct comparison to the experiment is however impossible, as we can only simulate the early stages of precipitation, before the coarsening stage (<6,000 h), while the data reported in the experiment fall well within the coarsening stage (50,000-90,000 h). The simulation of the coarsening stage is computationally prohibitive due to the mechanisms acting in FeCu alloys: the vacancy is typically trapped in a Cu-rich precipitate for many AKMC steps leading to the migration of the precipitate itself as a whole [58]; coarsening then occurs by merging precipitates. To simulate this mechanism in a computationally efficient way a hybrid atomistic/object kinetic Monte Carlo is necessary, as applied for example in [66]. The definition of the parameters for such simulations in a multicomponent system, however, requires the production of a large amount of precipitate mobility data, well beyond the scope of the present work. We therefore only focus on the nucleation and growth stages under thermal ageing. Note that for the composition given in Table 7, at 638 K all elements except Cu are fully soluble when in the binary sub-systems (see Table 2 and Figure *2*).

The values for $E_{m0}$ used in the KW approximation (see equation 6) are taken from [56] and summarized in Table 7. The vacancy formation energy used for the time rescaling (see equation 5) was chosen to be 1.8 eV (see Table 4).

To understand the effect of the different alloying elements, they were added incrementally. Starting from the FeCu alloy, in which precipitation necessarily occurs under the given thermal conditions, we address the effects of adding Mn and Ni, studying FeCuMn, FeCuNi and FeCuMnNi alloys. Next, taking FeCuMnNi as baseline alloy, we investigate the effects of the addition of traces of Cr, Si and P. We thus investigate the FeCuMnNiCr, FeCuMnNiSi and FeCuMnNiP alloys. Finally, the complete FeCuMnNiCrSiP alloy was simulated.

Table 7 – Composition and $E_{m0}$ used in the simulations.

|  | Fe | Cu | Ni | Mn | Cr | P | Si |
|---|---|---|---|---|---|---|---|
| Composition (at.%) | balance | 0.44 | 1.66 | 1.38 | 0.054 | 0.018 | 0.75 |
| $E_{m0}$ (eV) | 0.70 | 0.51 | 0.63 | 0.42 | 0.53 | 0.40 | 0.51 |

For each alloy composition, six independent ageing experiments were simulated to allow for statistics. The error bar on number density and average size of the solute clusters was estimated as twice the standard error, which corresponds to the 95% confidence interval. The width of the curves presented in Figure 6 and Figure 7 corresponds to the latter error bar. The cluster identification is based on a first nearest neighbor criterion and the minimum cluster size considered in the statistics was 8 solutes.

The effect of Mn and Ni on the precipitation kinetics of FeCu is shown in Figure 6, in terms of average cluster density and size until the onset of coarsening. The addition of Ni and Mn delays the appearance of the precipitation peak, after which coarsening occurs, by about a factor 6 and 16, respectively, while the peak density increases by about a factor 4 and 2, respectively. The latter suggests that both Ni and Mn favor the nucleation of Cu precipitates, consistently with simulations in [57, 86]. Specifically, the presence of either Ni and Mn leads to the formation of several small, but stable, clusters, thereby leading to a higher final density, but smaller size, of the Cu-rich precipitates. However, there is a qualitative difference between Ni and Mn in terms of kinetics: while the former seems to accelerate the onset of precipitation (nucleation), Mn delays it. For the combined effect of Ni and Mn, we observe that the maximum density is similar to the FeNiCu case while the delay in the appearance of the peak is similar to the FeMnCu case, although the onset of precipitation is somewhat accelerated. Thus, for the FeNiMnCu alloy, the delayed appearance of the peak is mainly determined by Mn, while the increased cluster density and faster onset is mainly due to Ni. Thus, compared to FeCu, the addition of Mn and/or Ni delays the onset of coarsening.

The effect of Cr, Si and P on the precipitation kinetics in FeNiMnCu is presented in Figure 7. We observe that the Cr content has no impact, while P introduces a delay in peak density of about a factor 10 compared to FeNiMnCu, without effecting its height. In contrast, the presence of Si increases the peak density by about a factor 2, while delaying it by about a factor 40, compared to the FeNiMnCu alloy. When Cr, Si and P are simultaneously present, the same evolution as in the FeNiMnCuSi alloy is followed. Thus, in the latter alloy the effect of Si dominates, most likely due to the significantly higher concentration, as compared to the other two elements.

Globally, compared to the FeCu binary, the combination of Mn, Ni, Si, P and Cr delays the onset of coarsening by a factor 600-700 and increases the peak density by a factor 10. Except for Cr, all elements have thus a strong impact on the precipitation kinetics.

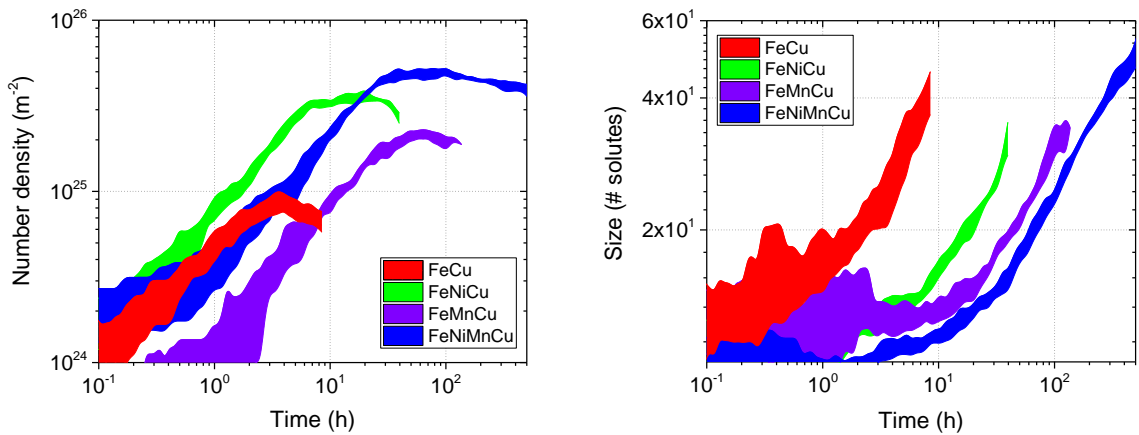

Figure 6 – Number density and size of the observed solute clusters as a function of ageing time: FeCu is the baseline case, on which the effect of adding Ni, Mn and both is evaluated.

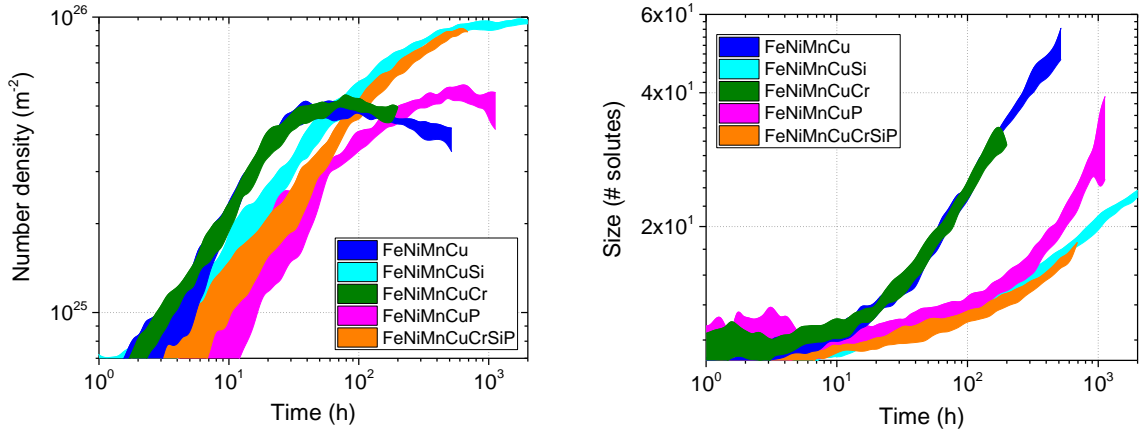

Figure 7 – Number density and size of the observed solute clusters as a function of ageing time: FeNiMnCu is the baseline case, on which the effect of adding Si, Cr, P and their combination is evaluated.

Next, we investigate the type of clusters that form depending on the alloy composition. For each alloy, we analyzed the cluster population of the last frame for all six independent runs, thereby collecting statistics for each alloy up to 180-890 solute clusters. Figure 8 and Figure 9, provide the histogram (normalized to unity) of the composition of the solute clusters for all investigated alloys. Examples of atomic configuration of solute clusters are also shown in the figures. We note that even though the cluster detection algorithm cannot detect Fe atoms in the clusters, we observed by visual inspection that in none of the clusters Fe atoms were present near their center.

In the FeCu alloy, only pure Cu clusters were observed, consistently with the simulations performed in [66] and with thermodynamic expectations based on a large number of experiments (see for example [87-90]). By adding Mn, a single population of mixed MnCu clusters with mean composition ~60% Mn and ~40% Cu is observed. Similarly, the addition of Ni leads to the formation of a single population of mixed CuNi clusters with mean composition ~35% Cu and ~65% Ni. In both cases the precipitates are richer in either Mn or Ni than Cu. When Mn and Ni are simultaneously added, two distinct populations of solute clusters appear: NiCu and MnNi clusters. As illustrated in Figure 8, these clusters may merge together into larger clusters, but with spatial separation between both cluster types. This view is consistent with the histograms shown in Figure 8, which indicates that clusters with and without Cu are formed. Although not clear from the histogram, we observed that no MnCu clusters are formed. We note that the two populations of Mn containing clusters correspond to MnNi clusters and merged MnNi/NiCu clusters, not MnCu clusters. This observation is consistent with the higher $E_\text{b}$ of MnNi pairs compared to MnCu pairs (see Figure 1). From the histogram we observe that the Ni and Cu content is similar for all clusters, 40-55% and 0-20%, respectively. The Mn content, on the other hand, spans a wider range 20-70%.

Taking now the FeNiMnCu alloy as the reference, Cr does not essentially change the solute cluster composition and morphology, and is sporadically found in the periphery of solute clusters. This observation is consistent with repulsion or neutral interaction with the other solutes (see Figure 1 and Table 1). The addition of P also does not essentially change the solute cluster composition or morphology. Due to the small P content in the matrix, little P is found in solute clusters, however, P binds quite strongly with Ni, Mn and Cu (see Figure 1 and Table 1) and is therefore found at the center of the solute clusters. In conclusion, for the investigated small concentrations, neither Cr nor P change the morphology or composition of MnNiCu clusters.

The addition of Si, in contrast, significantly changes the solute cluster composition and morphology. Because of the presence of Si, all solutes bind together in a single cluster thereby removing the spatial separation between NiCu and MnNi clusters. Consequently, we observe SiMnNiCu clusters in a broad composition range, as indicated by "flat distributions" in the histograms. We note that the Si content of the cluster is at most 45% and thus still in the validity range of the CE, as discussed in section 3.2. Finally, the combination of Cr, P and Si together follows the results of the addition of Si. Only few SiMnNiCu clusters contain a Cr or P atom located at the periphery of the cluster. The fact that the addition of Si expels P to the periphery of the cluster can be explained by the SiP repulsion (see Figure 1): at the periphery of the cluster, the number of SiP bonds is minimal. Most clusters do not contain P or Cr, which is likely due to their small concentrations.

Thus, we can conclude that Si plays an essential role in mixing the solutes homogeneously into a single solute cluster, i.e., Si removes the spatial separation between Cu and Mn in MnNiCu clusters.

Figure 8 – Histogram of the composition of clusters for ternary and quaternary alloys. Representative examples for the clusters' morphology are added. The atoms are represented as: ● – Cu; ● – Ni; ● – Mn.

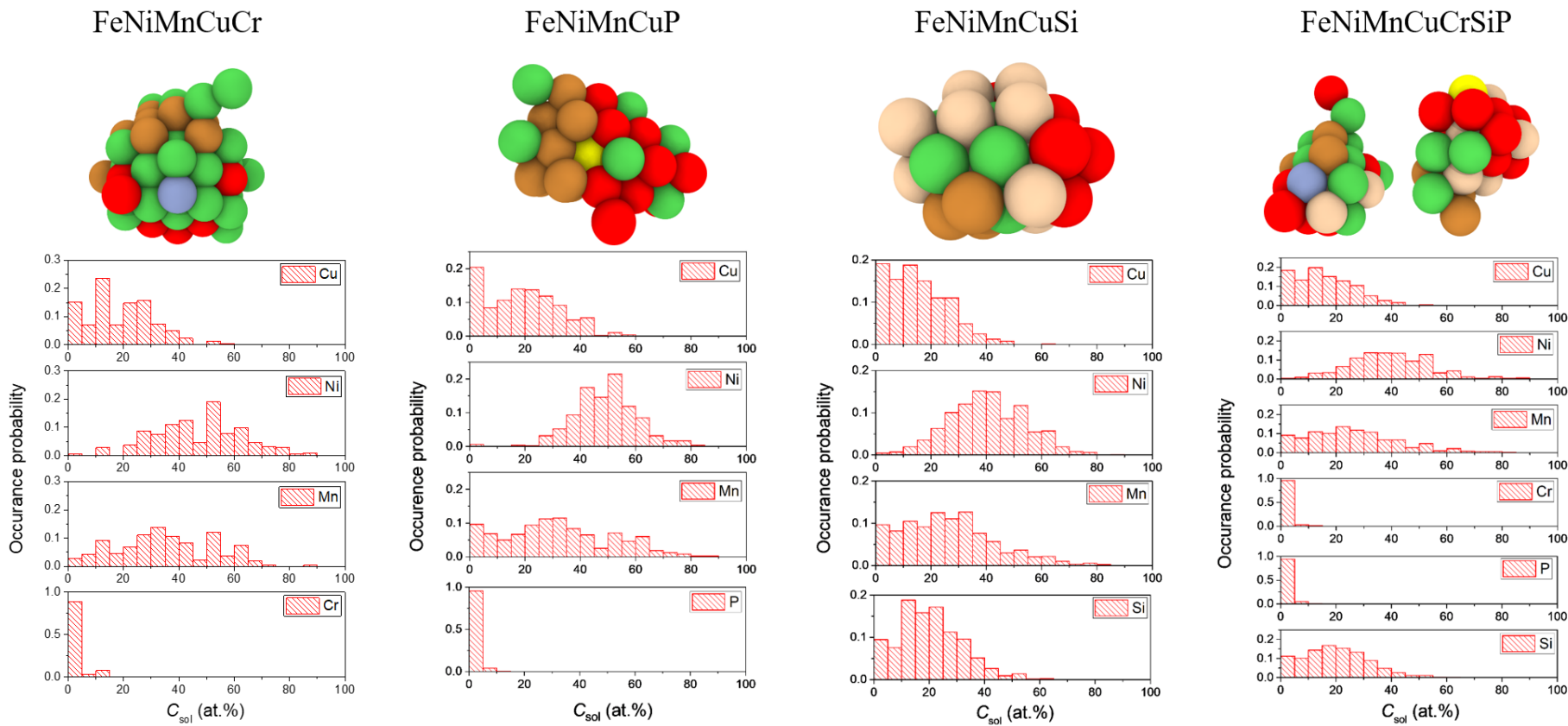

Figure 9 – Histogram of the composition of clusters for ternary and quaternary alloys. Representative examples for the clusters' morphology are added. The atoms are represented as: ● – Cu; ● – Ni; ● – Mn; ● – Cr; ● – P; ● – Si.

## 4. Conclusive remarks

We investigated the stability of small vacancy/solute clusters in the bcc Fe matrix by means of density functional theory (DFT). Over 700 configurations were studied, producing binding energy data that can be readily exploited in cluster dynamics and object kinetic Monte Carlo codes. The main conclusion is that, although CrCr, SiSi, PP and SiP pairs repel, they can be stabilized in larger clusters (starting from triplets) by the addition of other alloying elements, such as Mn, Ni and Cu.

Next, we fitted and validated a cluster expansion (CE) to a large DFT database, as well as to the experimental solubility of the different solutes in bcc Fe. We find that triplet many-body interactions are sufficient to describe v, Mn, Ni, Cu clusters but insufficient to describe Cr, Si and P clusters. Nevertheless, triplet many-body interactions included in the CE suffice if the clusters do not contain more than 75% of Cr, Si or P, thus this level of complexity may be considered sufficient.

The obtained CE can be readily used as an accurate energy model that extrapolates and interpolates the DFT data to describe vacancy/solute clusters in the bcc Fe matrix. We applied the CE to simulate the early stages of thermal ageing of an FeNiMnCuSiCrP alloy, representative a typical high-Cu/high-Ni RPV steel, and investigated the isolated effects of the alloying elements. We find that both Mn and Ni delay the onset of coarsening by about an order of magnitude compared to the FeCu binary. The addition of the minor alloying elements P/Si reduces the onset of coarsening by an additional order of magnitude. Mn, Ni and Si are altogether found to significantly increase the number density of the solute clusters. Si plays an essential role in the formation of spatially mixed MnNiCuSi cluster. Without Si, two cluster populations are formed, MnNi and CuNi clusters that merge into larger clusters whilst keeping the spatial separation between Ni- and Cu-rich parts.

The most striking result of the present study is the fact that a CE fitted accurately to a large number of DFT calculated solute-solute and solute-vacancy binding energy values, but without in any way biasing the Hamiltonian towards the stabilization of specific thermodynamic phases, does not spontaneously lead to the appearance of any known phase (for example B2 NiMn). This is qualitatively in agreement with the wide spectrum of possible phases and compositions that a recent in-depth DFT study revealed for the precipitates that may form in the chemically complex systems addressed in this work [91]. Since only the early stages of precipitation could be simulated in this work, without any coarsening and limited growth, the precipitates produced in the simulation are very small. For the sizes that could be sampled, the stabilization of any specific and well-identified phase, in a framework of a wide range of possible ordered structures and compositions, is clearly not possible. This suggests that under irradiation, where the presence of point-defects will not only change the kinetics of precipitation, but also influence the stability of possible clusters, the variety of compositions encountered will further multiply, making it difficult to associate the composition of the solute clusters that are formed and observed with defined and known thermodynamic phases. In order to explore such situations, as an outlook this CE will be used as a starting energy model in which radiation effects (essentially interstitials in the form of dumbbells) can be included.

**Acknowledgments**

This work was partly supported by the H2020 European project SOTERIA (No. 661913) and the FP7-Euratom-Fission European project MATISSE (No. 604862). It also contributes to the Joint Programme on Nuclear Materials of the European Energy Research Alliance.

**Appendix: Cluster expansion parameterization**

Any configuration in a crystalline system (or supercell) $\Omega$ containing $N$ lattice sites and $K$ constituents can be described by a vector $\boldsymbol{\sigma} = (\sigma_1, \ldots, \sigma_N)$ containing $N$ site operators $\sigma_n$ that, depending on the species occupying site $n$, take an integral value [92]. For the vFeCrMnNiCuSiP system with $K=8$, we chose $\sigma_n$ to take the values 0,1,2,3,4,5,6,7 if site $n$ is occupied by v, Fe, Cr, Mn, Ni, Cu, Si or P, respectively.

The configurational energy of the system with configuration $\boldsymbol{\sigma}$ can be expressed as [93, 94],

$$E(\boldsymbol{\sigma}) = J_0 + \sum_{\alpha \subseteq \Omega} \sum_{\boldsymbol{s}} m_\alpha J_\alpha^s \xi_\alpha^s(\boldsymbol{\sigma}), \tag{A1}$$

where the summation over $\alpha$ runs over all symmetrically distinct clusters that can be identified on the lattice $\Omega$ and the summation over $\boldsymbol{s} = (s_1, \ldots, s_{|\alpha|})$ runs over all possible combinations of $\boldsymbol{s}$. Here $s_i$ can take the values 1,…, $K$-1. Thus, $\boldsymbol{s}$ is a vector of integers (here $\boldsymbol{s} = (1,2,3,4,5,6,7)$), which define the specific type of configuration function associated with each site in the cluster $\alpha$. In equation A1, the energy is per atom, $J_\alpha^s$ are the effective cluster interactions (ECI), $\xi_\alpha^s(\boldsymbol{\sigma})$ the cluster correlation functions, and $m_\alpha$ is the multiplicity, which denotes the number of equivalent clusters $\alpha$ per lattice site.

When the sum over all possible combinations of $s$ is restricted to the non-equivalent ones, the multiplicity factor also takes the index $s$, accounting for the number of equivalent combinations. In practice, the sum over $\alpha$ is truncated to a maximum cluster $\alpha_{\max}$. Physically this means that all interactions are contained in such a $\alpha_{\max}$. In our case, we chose the triplet (see Figure 1 for the topology) as $\alpha_{\max}$.

The correlation functions mentioned in equation A1 are defined as,

$$\xi_\alpha^s(\boldsymbol{\sigma}) = \langle \prod_{i=1}^{|\alpha|} \phi_{s_i}(\sigma_i) \rangle_\alpha, \tag{A2}$$

where the average runs over all symmetry equivalent clusters in the lattice $\Omega$ and $\phi_{s_i}$ are the so-called configuration functions, which serve as basis functions to span the configurational space (see [93, 94] for more details). In this work we chose orthogonal configuration functions following [95, 96] described as,

$$\phi_{s_i}(\sigma_i) = \begin{cases} 1 & \text{if } s_i = 0 \\ -\cos\left(2\pi \left\lceil \frac{s_i}{2} \right\rceil \sigma_i / K\right) & \text{if } s_i > 0 \text{ and odd} \\ -\sin\left(2\pi \left\lceil \frac{s_i}{2} \right\rceil \sigma_i / K\right) & \text{if } s_i > 0 \text{ and even} \end{cases}, \tag{A3}$$

with $\lceil x \rceil$ a function that ceils $x$ to the nearest integer. These $K$ basis functions span the $K^N$ configurational space. The optimized coefficients for the CE are given in the supplementary material.